\documentclass[twocolumn,twoside,amsmath,amssymb,superscriptaddress,showpacs,nofootinbib,aps]{revtex4-1}
\usepackage{graphicx}
\usepackage{float}

\usepackage{slashed}
\usepackage{amsmath}
\usepackage{amsthm}
\usepackage{subfigure}
\usepackage[utf8]{inputenc}
\usepackage{graphicx}
\usepackage{epsfig}
\usepackage{amssymb}
\usepackage{dcolumn}
\usepackage{bm}

\newcommand{\ba}{\begin{array}}
\newcommand{\ea}{\end{array}}
\def\br{\begin{eqnarray}}
\def\er{\end{eqnarray}}
\def\be{\begin{equation}}
\def\ee{\end{equation}}
\usepackage{hyperref}

\def\({\left(}
\def\){\right)}

\def\<{\left\langle}
\def\>{\right\rangle}

\begin{document}

\title{Exclusive production of pions and the pion distribution amplitude}

\author{D. A. Fagundes} 
\affiliation{Departamento de Ci\^encias Exatas e Educa\c{c}\~ao, Universidade Federal de Santa Catarina - Campus
Blumenau, 89065-300, Blumenau, SC, Brazil}
\author{E. G. S. Luna}
\affiliation{Instituto de F\'{\i}sica, Universidade Federal do Rio Grande do Sul, Caixa
Postal 15051, 91501-970, Porto Alegre, RS, Brazil}
\author{A. A. Natale}
\affiliation{Instituto de F\'{\i}sica Te\'orica, Universidade Estadual Paulista (UNESP), Rua Dr. Bento T.
Ferraz, 271, Bloco II, 01140-070, S\~ao Paulo - SP, Brazil }
\author{M. Pel\'aez}
\affiliation{Instituto de F\'{\i}sica, Universidade Federal do Rio Grande do Sul, Caixa
Postal 15051, 91501-970, Porto Alegre, RS, Brazil}
\affiliation{Instituto de F\'{\i}sica, Facultad  de  Ingenier\'{\i}a, Universidad  de  la  Rep\'ublica,
J.  H.  y  Reissig  565,  11000  Montevideo,  Uruguay}

%


\begin{abstract}

Considering, as a limit case, an approximately flat pion distribution amplitude, which is determined from the hardest, in momentum space, solution of the
Bethe-Salpeter equation for the pion wave function, we compute the pion transition form factor $F_{\pi\gamma\gamma^{*}}(Q^{2})$ and the pion form factor
$F_{\pi}(Q^2)$, taking into account the LO as well as NLO form of the hard coefficient function entering the leading-twist factorization formula. We also
compute the exclusive photoproduction of pions pairs at high energies, $\gamma \gamma \to \pi^{+}\pi^{-}$, where perturbative QCD can be used to compute
the hard scattering matrix elements. We verify that the existent data for exclusive pion production can be reasonably described as a function of such flat
distribution amplitude.

\end{abstract}
\date{\today}
\maketitle

\vskip 0.5cm                           

\section{Introduction}

\noindent

The hadronic distribution amplitudes (DAs) are an essential ingredient for measuring exclusive processes at large momentum transfer. Most of the recent
data about the Standard Model parameters rely on QCD factorization, asymptotic freedom and make use of process-independent hadronic DAs.
This fact reveals the importance of knowing the intricacies of quarks and gluons within the hadrons. The pion, being the simplest hadron, should, in
principle, be the easiest particle to offer a laboratory to learn about hadronic DAs, although its study is still motive of debates.

Some years ago the BaBar Collaboration \cite{babar} published results for the photon-pion transition form factor $F_{\pi\gamma\gamma^{\ast}}(Q^{2})$, where one
of the photons is near mass shell ($Q^{2}\approx0$) and the other one is far off mass shell (large $Q^{2}$).
These measurements have been taken in single-tagged two-photon $e^{+}e^{-}\rightarrow e^{+}e^{-}\pi^{0}$ reaction and have been performed in a wide range of
momentum transfer squared ($4-40$ GeV$^{2}$). It is expected that standard factorization approach can be applied at such high $Q^2$ region \cite{bl}.

The surprise with the BaBar result is that it was not in agreement with the expected perturbative QCD behavior, where
$Q^{2}F_{\pi\gamma\gamma^{\ast}}(Q^{2}\rightarrow\infty)$ should be limited to the value $\sqrt{2}f_{\pi}\approx0.185$ GeV, which is known as the BL limit
\cite{brodsky2}. Here $f_\pi = 131$ MeV is the pion decay constant.
Some time later the Belle Collaboration presented data \cite{belle} in the same range of transferred momenta showing that the pion transition form factor
may not increase as fast as indicated by the BaBar results, although some medium values of Belle data also appear to be in contradiction with the BL limit. 

These experiments originated several theoretical papers speculating why the data should or should not obey the BL limit
\cite{rad,poly,te1,te2,te3,te4,te5,te6,te7,te8,te9}. Some of these and recent proposals also claimed that the pion distribution amplitude (DA) at high
momentum transfer was not given by the asymptotic form \cite{efr}
\begin{equation}
\varphi_{\pi}^{as}(x)=6x(1-x) ,
\label{eq0}
\end{equation}
but should be replaced by a broad concave distribution \cite{roberts001,roberts002} or a flatter one \cite{rad,poly,te2,te6,f1,f2,f3,f3l}. Available
information indicates that the above asymptotic distribution is a poor approximation to the pion distribution amplitude even at large momentum scales
\cite{roberts001}. As a consequence, predictions of leading-order, leading-twist formula based on $\varphi_{\pi}^{as}(x)$ should be revisited.
Actually, a flat DA is consistent with the BaBar data \cite{rad}, although a theoretical support for such possibility is still missing. Thus we may assume,
as claimed in Ref.\cite{huang}, that there is no definite conclusion as yet on which is the asymptotic form of the pion DA, and it is possible that a
combined analysis of data of the processes involving pions would shed light on the pion DA \cite{huang2}.

The pion transition form factor is quite dependent on the pion distribution amplitude, and this one is directly related to the pion wave function.
Recently some of us proposed a limit on the transition form factor based on the hardest solution (in momentum space) of the Bethe-Salpeter equation (BSE)
for the pseudoscalar pion state \cite{luna}. This wave function leads to the flattest QCD DA and such kind of behavior, as argued by Radyushkin
\cite{rad}, can describe the BaBar data. The pion DA obtained in \cite{luna} shows that non-perturbative effects change the soft asymptotic behavior of
the pion wave function leading to a much broader DA than the one of Eq. (\ref{eq0}), and this fact was observed in lattice simulations \cite{latchin}.
Therefore a very flat (not constant) pion DA can be naturally explained within the QCD theory when associated to a particular behavior of the BSE solution.

In this work we will explore in detail the predictions of this extreme BSE solution for the high energy behavior of the pion transition form factor, its
form factor and the two-photon production of a pion pair. The paper is organized as follows: in Section II we first illustrate our theoretical framework
\cite{luna} and recall how the pion distribution amplitude can be obtained from the BSE. We also advocate in favor of a BSE solution for the pion wave
function that decreases slowly with the momentum, which is at the origin of the flat pion DA. In Section III the pion DA introduced in the previous section
is used to determine the pion transition form factor. In Sections IV and V we continue with the phenomenological implications of our flat DA respectively
in the cases of the pion form factor and hard exclusive two-photon production of a pion pair. Section VI contains our conclusions.

\section{The pion distribution amplitude from the BSE}

We start by writing down the pion distribution amplitude at leading twist, as a function of the pion-quark vertex and the quark self-energy \cite{luna,dor}:
\begin{align}
\varphi_{\pi}(x)  &  =\frac{N_{c}}{4\pi^{2}f_{\pi}^{2}}\int_{-\infty}^{\infty
}\frac{d\lambda}{2\pi}\int_{0}^{\infty}du\frac{F(u+i\lambda\bar{x},u-i\lambda
x)}{D(u-i\lambda x)D(u+i\lambda\bar{x})}\nonumber\\
&  \times\left[  x\Sigma(u+i\lambda\bar{x})+\bar{x}\Sigma(u-i\lambda
x)\right]  \,, \label{eq1}%
\end{align}
where $\bar{x}=(1-x)$, the variable $u$ represents the quark transverse momentum squared, $\lambda x$ and $-\lambda\bar{x}$ are the longitudinal
projections of the quark momentum on the light cone directions, and $\Sigma(u)$ is the dynamical quark mass. The function $D(u)$ in the expression above
is a function related to the quark propagator,
\begin{equation}
D(u)\equiv u+\Sigma^{2}(u), \label{Propagator}%
\end{equation}
whereas $F$ is the momentum dependent part of the quark-pion vertex. The function $F$ can be approximated by
\begin{equation}
F(p^{2},{p^{\prime}}^{2})=\sqrt{\Sigma
(p^{2})\Sigma({p^{\prime}}^{2})}, \label{Propagator001}%
\end{equation}
where $p$ and ${p^{\prime}}$ are the quark and anti-quark momenta, respectively. The pion DA at leading twist is usually normalized as
\begin{equation}
\int_{0}^{1}dx\,\varphi_{\pi}(x,\mu)=1 , \label{eq1b}%
\end{equation}
where $\varphi_{\pi}(x,\mu)$ is defined at some normalization scale $\mu$.

As demonstrated by Delbourgo and Scadron some years ago \cite{ds}, the spontaneous generation of fermion mass can be associated with zero-mass pseudoscalar
bosons. Specifically, the Schwinger-Dyson equation for the dynamical quark self-energy, $\Sigma_{SD}(p^{2})$, is identical to the BSE for a pseudoscalar,
$\Phi_{BS}^{P}(p,q)$, at zero momentum transfer:
\begin{equation}
\Sigma_{SD}(p^{2})\approx\Phi_{BS}^{P}(p,q)|_{q\rightarrow0}. \label{eq00}%
\end{equation}
This relation is a consequence of the fact that $\Sigma$ and $\Phi$ are related through the Ward-Takahashi identity.
The homogeneous BSE can be written as \cite{BS001}
\begin{equation}
\Phi(k,P)= -i\int_{q}^{\infty}\frac{d^{4}q}{(2\pi)^{4}}\, K(k;q,P)\,
S(q_{+})\, \Phi(q;P)\, S(q_{-})\,, \label{eq02}%
\end{equation}
where $P$ ($q$) is the total (relative) momentum of the quarks, $K$ is the fully amputated quark-antiquark scattering kernel, $S(q_{i})$ are the dressed
quark propagators, $q_{-}=q-(1-\eta)P$, and $q_{+}=q+\eta P$, with $0\leq\eta\leq1$. Here $\eta$ is the momentum fraction parameter. In the pion case the
homogeneous BSE is valid on-shell, i.e. $P^{2}=0$. Note that in Eq. (\ref{eq02}) we have suppressed all indices (color, etc...).

The quark masses are dynamically generated along with bound state Goldstone bosons (the pions). It is worth noting that the Eq. (\ref{eq02}) is an integral
equation that can be transformed into a differential equation of second order. The two solutions of this differential equation, as obtained in Refs.
\cite{lane,lan}, are characterized by one soft asymptotic solution
\begin{equation}
\Phi_{\pi}^{R}(p^{2})\sim\Sigma^{R}(p^{2} \gg m_{q}^{2})\sim\frac{m_{q}^{3}}{p^{2}%
}\,, \label{eq03}%
\end{equation}
and by an extreme hard asymptotic behavior of a bound-state wave function
\begin{equation}
\Phi_{\pi}^{I}(p^{2})\sim\Sigma^{I}(p^{2} \gg m_{q}^{2})\sim m_{q}\left[
1 + b g^{2}  \ln\left( \frac{p^{2}}{m_{q}^{2}} \right)  \right]
^{-\delta}\,, \label{eq1a}%
\end{equation}
where $b=(11N_{c}-2n_{f})/48\pi^{2}$ is the first coefficient of the perturbative $\beta$ function (with $N_{c}=3$), $c=4/3$ is the Casimir eigenvalue for
quarks in the fundamental representation, $\delta = c/2b$, $g^{2}$ is the coupling constant, and $m_{q}$ is the dynamical quark mass at zero momentum.

The asymptotic expression shown in Eq. (\ref{eq1a}) satisfies the Callan-Symanzik equation, and was determined in the appendix of Ref. \cite{cs}.
The hard solution is constrained by the BSE normalization condition \cite{man}, which imply the condition $n_{f}>5$ \cite{lane,us3} (for this reason we
adopt in this paper $n_{f}=6$).
Note that the hard expression
(\ref{eq1a}) is an alternative solution to the soft one ($\Sigma(p^{2})\sim1/p^{2}$) \cite{poli}, which in turn leads to the standard distribution amplitude
$\varphi_{\pi}^{as}(x,\mu \to \infty)=6x(1-x)$.
It is known that we may have solutions with a momentum behavior varying between the equations (\ref{eq03}) and (\ref{eq1a}). The effective behavior of the
solution, particularly as the number of fermions is increased, depends on the theory dynamics \cite{us3,takeuchi}.

It has been argued that, in a scenario where the gluons have a dynamically generated mass and the chiral symmetry breaking is associated to
confinement, the solution (\ref{eq1a}) may be a realistic one \cite{us3,us1,us2}. The hard solution also appears associated to a finite quark condensate
when using the technique of the improved renormalization group approach in QCD \cite{chan}, and the condensate minimizes the vacuum energy provided that
$n_{f}>5$ \cite{mon}. Furthermore, it is well known that the hard solution is the only one consistent with an expansion group (Regge-pole like) solution
\cite{lan}.

Recently it was demonstrated numerically \cite{natale001} and analytically \cite{natale002} that (\ref{eq1a}) emerges when the current quark masses are
generated dynamically, although in these cases the power $\delta$ will depend on the details of the model. It is interesting to recall that
models with origin in the Nambu-Jona-Lasinio model (NJL), like the ones of Ref.\cite{f3,f3l}, also describe the data with a flat pion DA, what is not
surprising since the NJL model naturally lead to dynamical masses with a behavior similar to the one of Eq. (\ref{eq1a}), which, as shown in the sequence,
induce a quite flat DA.

The important fact to be noticed here is that Eq. (\ref{eq1a}) gives the hardest asymptotic behavior (in momentum space) allowed for a bound state solution
in a non-Abelian gauge theory. This behavior has the consequence that, no matter the hard solution is realized in Nature or not, the pion DA will be very
flat. Note that in our case a totally flat (constant) DA is not allowed since such a behavior can only be related to a fundamental pion. Thus any other
flatter pion DA than the one obtained from the hard solution (\ref{eq1a}) cannot be a realistic BSE wave function, and consequently would not be
consistent with a composite pion.

We compute the pion DA performing an integral over the full range of momenta (up to $p^2 \rightarrow \infty $), covering all possible
thresholds. In order to explore the full behavior of the ``hardest'' quark self-energy, we adopt a simple interpolating expression for
$\Sigma(p^{2})$ \cite{us3,us2}:
\begin{equation}
\Sigma(p^{2}) = m_{q} \left[  1+bg^{2} \ln\left(
\frac{p^{2} + m_{q}^{2}}{m_{q}^{2}}\right)  \right]  ^{-\delta}\,. \label{eq2}%
\end{equation}

Note that the Eq. (10) assumes a constant IR behavior
for the quark self-energy. This is totally consistent with numerical solutions of the DSE, and the value of the quark mass does not impact strongly
on the results. However the gluon mass that act as an IR cutoff introduces some effect in the calculation of the perturbative matrix elements. Of
course, we should stick to values consistent with most phenomenological calculations of this quantity.

Before proceeding it may be worth emphasizing that the $m_{q}$ factors introduced into the logarithm term leads to
the right infrared behavior, namely $\Sigma(p^{2}\rightarrow0)=m_{q}$. Also, the coupling $g^{2}$, calculated at the chiral symmetry breaking scale
$\Lambda^{\prime}$, may be written as 
\begin{equation}
{g}^{2}(p^{2})=\frac{1}{b\ln[(p^{2}+4M_{g}^{2})/\Lambda_{QCD}^{2}]}\,,
\label{eq6}%
\end{equation}
where $\Lambda_{QCD}$ is the QCD characteristic scale and $M_{g}$ is an effective dynamical gluon mass \cite{cornwall}. The coupling is infrared finite,
with an value $M_g(0)\approx 2 \Lambda_{QCD}$, consistent with the phenomenological models of Ref. \cite{us3,us1,us2,us4,ap,ap1}.

The pion DA numerical result calculated with Eqs. (\ref{eq1}) and (\ref{eq2}), and constrained by Eq. (\ref{eq1b}), can be quite well reproduced by the
normalized form \cite{luna}
\begin{equation}
\varphi_{\pi}(x;\epsilon)=\frac{\Gamma\left(  2+2\epsilon\right)  }{\Gamma^{2}\left(
1+\epsilon\right)  }\,x^{\epsilon}(1-x)^{\epsilon}\,, \label{eq3A}%
\end{equation}
where
\be
\epsilon\approx0.0248, 
\label{epi}
\ee
which will be used in the following calculations. Note that, according to Radyushkin \cite{rad}, QCD corrections will barely affect such flat distribution
amplitude, where no dependence with the factorization scale will be assumed.

\section{Pion transition form factor}

At sufficiently high $Q^{2}$ it is expected that the standard factorization approach can be applied \cite{che,lep} (for a review, see \cite{bl}), and the
pion transition form factor is given by
\begin{align}
F_{\pi\gamma\gamma^{*}}(Q^{2}) = \frac{\sqrt{2}f_{\pi}}{3}\int_{0}^{1}dx \,
\varphi_{\pi}(x)T^{H}_{\gamma\pi}(x,Q^{2},\mu^\prime). \label{eqa}%
\end{align}
This equation is obtained assuming factorization of the pion distribution
amplitude $\varphi_{\pi}(x)$ and the hard scattering amplitude
$T^{H}_{\gamma\pi}(x,Q^{2},\mu^\prime)$ given by \cite{brodsky2,brodsky}%
\be
T^H_{\gamma\pi}(x,Q^{2},\mu^\prime) = T^H_{1}(x,Q^{2},\mu^\prime)+T^H_{2}(x,Q^{2},\mu^\prime),
\label{eqa1}
\ee
where $\bar{x}=1-x$, $x$ is the longitudinal momentum fraction carried by the quark in the meson and $\mu^\prime$ is an
arbitrary momentum scale which separates the hard and soft momenta regions. 

The hard-scattering amplitude $T^{H}_{\gamma\pi}(x,Q^{2},\mu^\prime)$ must be symmetrized under exchange $x \leftrightarrow \bar{x}$
\be
T^H_{2}(x,Q^{2},\mu^\prime) = T^H_{1}(\bar{x},Q^{2},\mu^\prime),
\ee
and at the next to leading order $T^{H}_1 (x,Q^{2},\mu^\prime)$ is given by \cite{aguila,braaten}
\begin{equation}
T^{H}_1 (x,Q^{2},\mu^\prime) = \frac{1}{xQ^{2}} \left\{ 1 + \frac{4}{3}\frac{\alpha_{s}(\mu^{\prime 2})}{2\pi} \times A(x,Q^2,\mu^\prime) \right\},
\label{eqb}
\end{equation}
where
\br
A(x,Q^2,\mu^\prime)&=&\left[ \frac{1}{2} \ln^{2} x 
- \frac{x \ln x}{2\bar{x}} \right.\nonumber \\
&-& \left. \frac{9}{2} + \left[ \frac{3}{2} + \ln x  \right]\ln \left( \frac{Q^{2}}{\mu^{\prime 2}} \right)  \right] .
\label{eqb1}
\er
For simplicity we set $\mu^\prime=Q$ and $T^{H}_1 (x,Q^{2},\mu^\prime)$ can be written as
\begin{equation}
T^{H}_1 (x,Q^{2}) = \frac{1}{xQ^{2}} \left[ 1 + \frac{4}{3}\frac{\alpha_{s}(Q^2)}{4\pi} f(x) \right] , 
\label{eqb0}
\end{equation}
where $f(x)$ is given by
\be
f(x)= \ln^2{x} - \frac{x\ln{x}}{\bar{x}}-9 .
\label{eqcor}
\ee

As emphasized by Radyushkin \cite{rad}, the finite size $R \approx 1/M$ of the
pion interaction should provide a cut-off for the $x$ integral. Therefore the $xQ^{2}$ in the
denominator of Eq.(\ref{eqb0}) should be changed as
\be
xQ^{2}\rightarrow xQ^{2} + M^2(xQ^2) \, .
\label{eqxx}
\ee
In principle the factor $M$ should be related to the dynamical quark mass. It was also proposed by
Radyushkin that $M$ could be treated as an effective gluon mass. Indeed the meson radius may have a deep connection with the
effective gluon mass as discussed in \cite{halnat}, and in the following we will assume $M(Q^2)\equiv M_g(Q^2)$.
Therefore, no matter we have one case or another, the asymptotic transition form factor will be given by
\be
  F_{\pi\gamma^{\ast}\gamma}\left(  0;Q^{2} \rightarrow\infty,0\right)  =\frac{\sqrt{2}}{3}f_{\pi}
	\int_{0}^{1}dx\frac{\varphi_{\pi}(x)}{xQ^{2}+ M_g^2}  .
\label{FAsMod}%
\ee
$M_g$, being a dynamical mass, should have a momentum dependence showing the decrease of the mass with the momentum.
However when  $xQ^{2}$ is small we can safely substitute
$M_g(xQ^2)$ by the infrared $M_g$ value in Eq.(\ref{FAsMod}), and for large $xQ^{2}$ the value of $M_g(xQ^2)$
is negligible compared to $xQ^{2}$. 

\begin{figure}[h]
\begin{centering}
\includegraphics[scale=0.45]{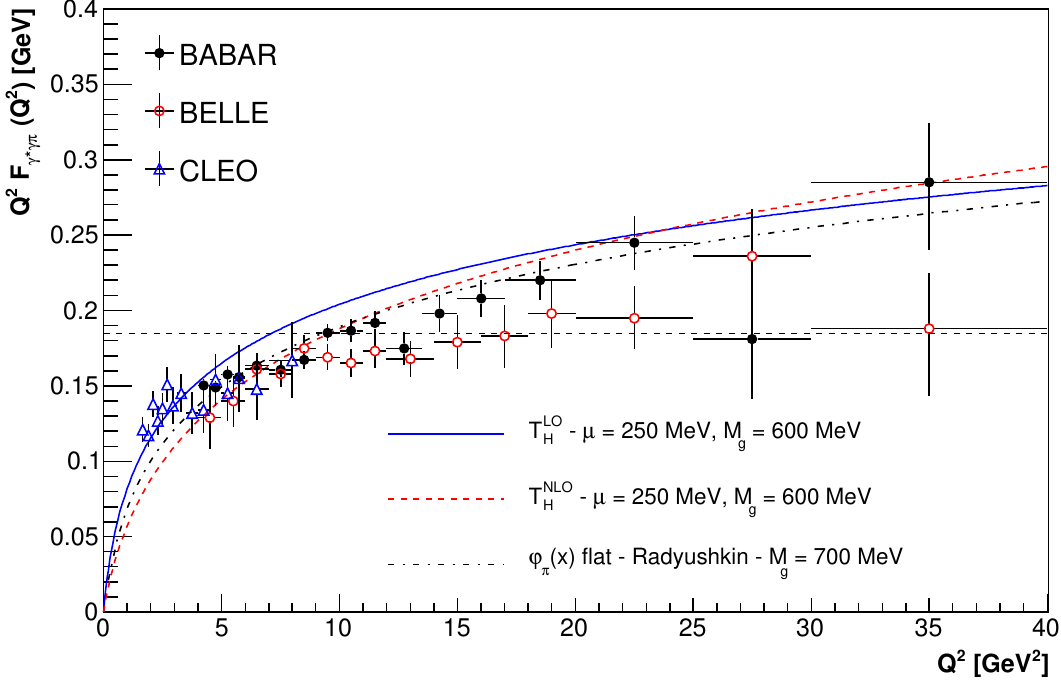} 
\par\end{centering}
\caption{Pion transition form factor calculated with the flat pion distribution of Eq.(\ref{eq3A}) considering dynamical
quark and gluon masses given respectively by $250$ and $600$ MeV. We also plot the Radyushkin result \cite{rad}
with a $700$ MeV gluon mass.}
\label{fig1} 
\end{figure}

Our result for the pion transition form factor, using Eq.(\ref{eq3A}) and the
hard-scattering amplitude at leading and next-to-leading order is shown in Fig.(\ref{fig1}), where it is possible to see
a reasonable agreement with the BaBar data. Note that the introduction of the NLO correction is important
for this agreement.

\section{The pion form factor}

\begin{figure}[h]
\begin{centering}
\includegraphics[scale=0.4]{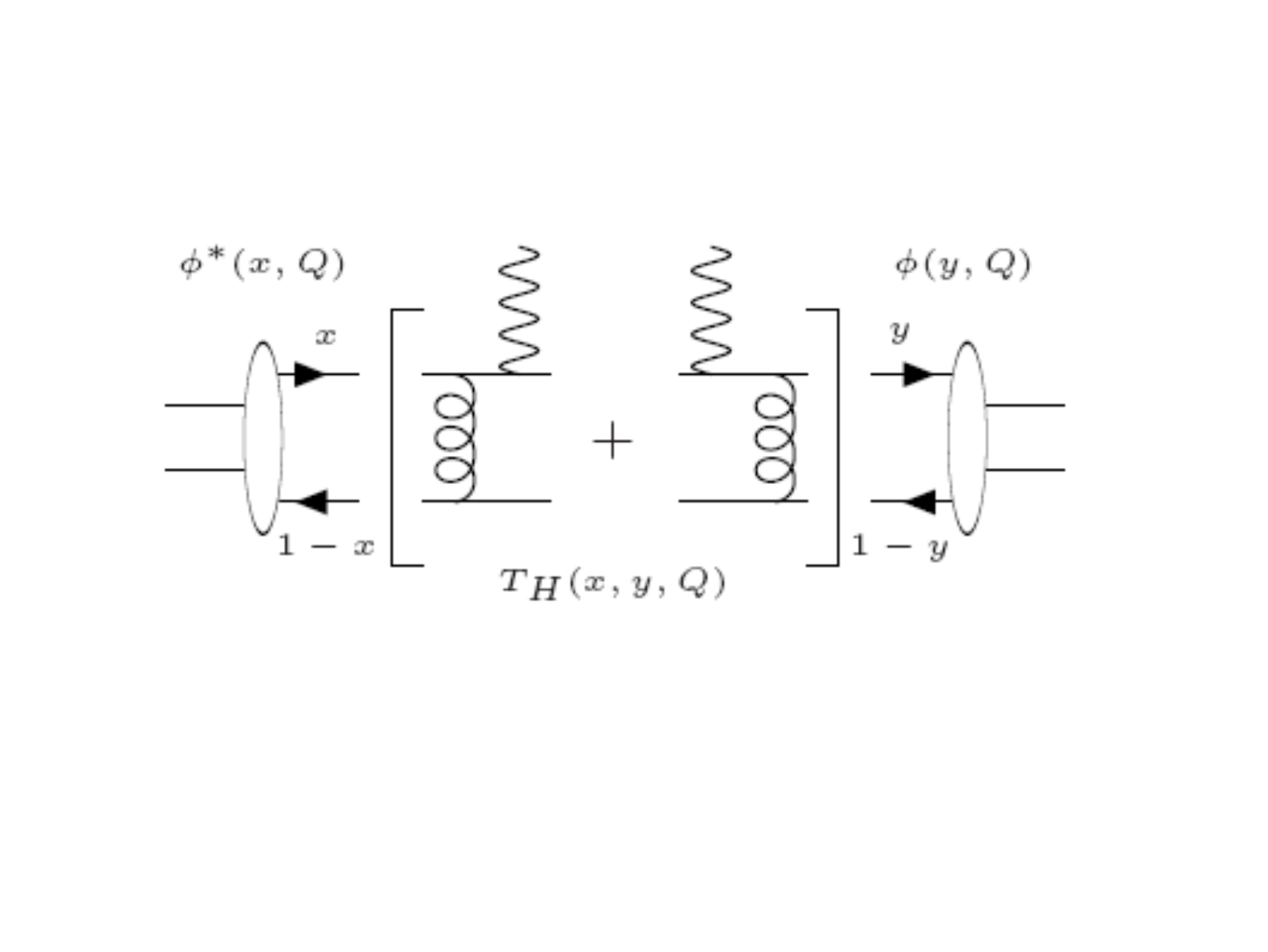} 
\par\end{centering}
\caption{The leading-order diagrams that contribute to
the pion form factor. $\phi(x,\tilde{Q}_x)$ is the pion wave
function, that gives the amplitude for finding the quark or
antiquark within the pion carrying the fractional momentum $x$ or
$1-x$. The photon transfers  the momentum $q^\prime$ (in Minkowski
space), $Q^2=-q^{\prime 2}$,  for the  $q\overline{q}$ pair of
total momentum $P$ producing a $q\overline{q}$ pair of final
momentum $P^{\prime}$.}
\label{PFF} 
\end{figure}

The pion form factor $F_{\pi}(Q^2)$ is also going to be changed if the pion DA is flatter than the
usual asymptotic form. As already discussed in Ref.\cite{aguinat} the QCD prediction for the
form factor is also dependent on the IR non-perturbative behavior of the gluon propagator and of the running
coupling constant \cite{ji}. Therefore we will now compute $F_{\pi}(Q^2)$ with the new DA discussed above
and also with improved non-perturbative results for the gluon propagator and coupling constant. 
The asymptotic form factor is predicted by
perturbative QCD \cite{bl,brodsky2,ji,brodsky3}. It depends on the internal pion
dynamics that is parametrized by the quark distribution amplitude
of the pion. The QCD expression for the pion form factor
is \cite{brodsky}
\br F_{\pi}(Q^2)= \frac{f_\pi^2}{12}\int_{0}^{1}\!\!dx && \! \int_{0}^{1}\!\!dy \,
\varphi^{*}
(y,\tilde{Q}_y)  \nonumber\\
&&\times T_H(x,y,Q^2)\varphi(x,\tilde{Q}_x),
 \label{fpi} \er
where $\tilde{Q}_x= \textnormal{Min}(x,1-x)Q$ and Q is the 4-momentum in Euclidean space transferred by the photon . The
function $\varphi(x,\tilde{Q}_x)$ is the momentum dependent pion DA,
that gives the amplitude for finding the quark or antiquark
within the pion carrying the fractional momentum $x$ or
$1-x$, respectively. 
$ T_H (x,y,Q^2)$ is the hard-scattering
amplitude that is obtained by computing the quark-photon
scattering diagram as shown in Fig.{\ref{PFF}}.

How the pion distribution amplitude (12) evolves with the scale ``$\tilde{Q}$'' will be discussed in Section VI. We advance
that independently of its shape at some specific low normalization point $\mu_{0} \lesssim 1$ GeV, at large values of $\mu$
the pion DA acquires an expected QCD asymptotic form. Furthermore, it is known that nonperturbative lattice calculations for
small normalizations scales $\mu \sim 0.5$ GeV \cite{dalley001} produced a rather flat DA very close to the one obtained is
this work. As we will see, at low scales dominates an extremely slow
non-perturbative evolution. Thus, even a choice $\mu = \tilde{Q}$ does not alter our results.

The lowest-order
expression of $ T_H(x,y,Q^2)$ is given by \cite{aguinat}
\br T_H(x,y,Q^2) &=& \frac{64\pi}{3}\left[
\frac{2}{3}\alpha_s(K^2)D(K^2) \right.\nonumber\\
& & \left. + \frac{1}{3}\alpha_s(P^2)D(P^2)\right].
\label{thDAC}
\er
where $K^2=(1-x)(1-y)Q^2$ and $P^2=xyQ^2$. Here $D(K^2)$ is related
to the perturbative QCD gluon propagator
that, in the Landau gauge, is given by

\be D_{\mu\nu}(q^2)= \({\delta}_{\mu\nu}
-\frac{q_{\mu}q_{\nu}}{q^2}\)D(q^2), \quad D(q^2)=\frac{1}{q^2} .
\label{landau} \ee
In our analysis the perturbative $D(q^2)=\frac{1}{q^2}$ is now substituted by the non-perturbative (in Euclidean space) expression
\be
D(q^2)=\frac{1}{q^2 + M_{g}^2(q^2)} ,
\label{propcorn}
\ee
where $M_g(q^2)$ is the dynamical gluon mass which is roughly given
by \cite{agui,papa} $M_g^2 (q^2) \approx {M_g^4}/{(q^2+M_g^2)}$. Since this mass
decays very fast with the momentum our calculations are not affected if we just
assume $M_g^2(q^2)\approx M_g^2$, as we took for granted in the previous section. 

The inclusion of radiative corrections in the hard-scattering
amplitude imply that $T_H(x,y,Q^2)$ has to be multiplied by
the factor \cite{aguila}
\be
[1-\frac{5}{6}\frac{\alpha_s(Q^2)}{\pi}] \, .
\ee
Note that in our calculations we are including the radiative corrections in the hard-scattering amplitude,
and assume that factorization happens at a scale $Q^2 > 1$ GeV$^2$.

\begin{figure}[h]
\begin{centering}
\includegraphics[scale=0.45]{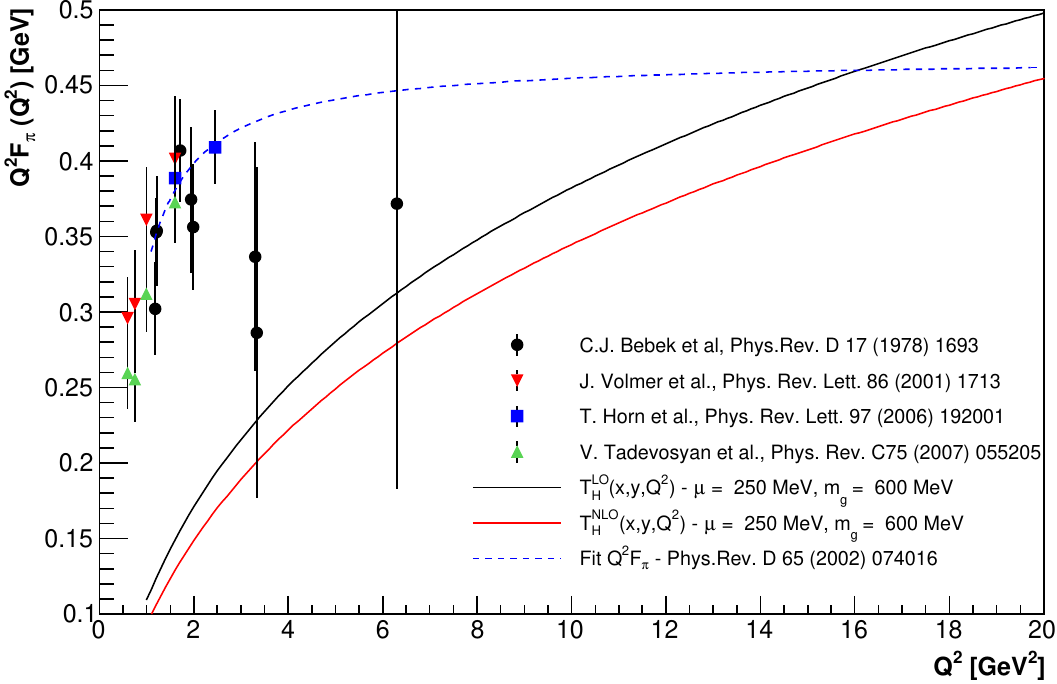} 
\par\end{centering}
\caption{Pion form factor calculated with the flat pion distribution of Eq.(\ref{eq3A}), and with dynamical
quark and gluon masses given respectively by $250$ and $600$ MeV. Comparison with the experimental fit of
Ref.\cite{yeh}. The experimental data are taken from \cite{bebek01,volmer01,horn01,tadevosyan01}.}
\label{fig2} 
\end{figure}

The result for the electromagnetic pion form factor is shown in Fig.(\ref{fig2}),
where it is compared to a simple fit to the experimental data \cite{yeh}:
\be
F_\pi^{fit}(Q^2)= \frac{0.46895}{Q^2} \left(1-\frac{0.3009}{Q^2}\right),
\label{eqfit}
\ee
although this is a quite naive fit, which does not include one of the highest energy data. It
is clear that more data is necessary in order to check the high energy behavior of the pion form factor, but
it is quite interesting that the high energy behavior of the electromagnetic form factor seems
to be reasonably described by the same factors (pion DA and dynamical masses) that we considered previously.
We observe that the pion form factor is not very
sensitive to $m_{q}$ and $m_{g}$, changing by about 15\% (19\%) when $m_{q}$ ($m_{g}$) ranges from 200 to 250 MeV (from 500 to 700 MeV).

\section{Hard exclusive two photon production of a pion pair}

The helicity amplitudes for a pion pair production in exclusive two photon collisions at high energies and
large center of mass scattering angles $\theta_{cm}$ is given by
\be
{\cal{M}}^{\lambda \lambda^\prime}=\int_0^1 dx \int_0^1 dy \varphi^*(x,\tilde{Q}_x)\varphi^*(y,\tilde{Q}_y)T^{\lambda \lambda^\prime}_H (x,y,Q^2),
\label{e21}
\ee
where ${\tilde{Q}}_x = \textnormal{Min}(x,1-x)\sqrt{s}|\sin{\theta_{cm}}|$, similarly for ${\tilde{Q}}_y$, and $s=W^{2}_{\gamma\gamma}$ is the square of the cm energy of
the two-photon system. $ T^{\lambda \lambda^\prime}_H (x,y,Q^2)$
is the helicity dependent perturbative hard scattering amplitude for two pion production.
The spin-averaged cross section for producing the pion pair is
\be
\frac{d\sigma}{dz}= \frac{1}{32 \pi s} \langle |\mathcal{M}|^{2} \rangle,
\label{e22}
\ee
with
\be
\langle |\mathcal{M}|^{2} \rangle = \frac{1}{4}{\sum_{\lambda \lambda^\prime}} \left| {\cal{M}}^{\lambda \lambda^\prime}\right|^2.
\ee
and $z=\cos \theta_{cm}$.
The hard scattering amplitudes (in leading order) for the different helicity structures are given by \cite{brodsky}
\br
\left. \ba{c}  { T^{(0)}_H (++) }\\ { T^{(0)}_H (--)}  \ea \right\} &=& \frac{16\pi \alpha_s}{3s} \frac{32\pi \alpha}{x(1-x)y(1-y)} \nonumber \\
&\times & \left[ \frac{(e_1-e_2)^2a}{1-z^2}\right] ,
\label{e23}
\er
\br
\left. \ba{c} { T^{(0)}_H (+-)} \\ { T^{(0)}_H (-+)} \ea \right\} &=& \frac{16\pi \alpha_s}{3s}  \nonumber \\
&\times &\frac{32\pi \alpha}{x(1-x)y(1-y)}\left[ \frac{(e_1-e_2)^2a}{1-z^2} \right. \nonumber \\
&+&  \frac{e_1e_2[x(1-x)+y(1-y)]}{a^2-b^2z^2} \nonumber \\
&+& \left. \frac{(e_1^2-e^2_2)(x-y)}{2}\right] ,
\label{e24}
\er
where $e_i$ are the quark charges (meaning that the pions have charges $\pm(e_{1}-e_{2})$) and
\be
\left. \ba{c} {a}\\ {b} \ea \right\} =(1-x)(1-y)\pm xy .
\label{e25}
\ee

In order to restrain the calculation at the perturbative QCD level we can multiply the right side of Eq.(\ref{e21}) by the
following form factor, which smoothly switches off the pQCD contribution
at low energies \cite{gs}
\be
F^{pQCD}(s)= 1-exp \left(\frac{-(s-4m_\pi^2)^4}{\Lambda^8_{pQCD}}\right) .
\label{e26}
\ee

\begin{figure}[h]
\begin{centering}
\includegraphics[scale=0.4]{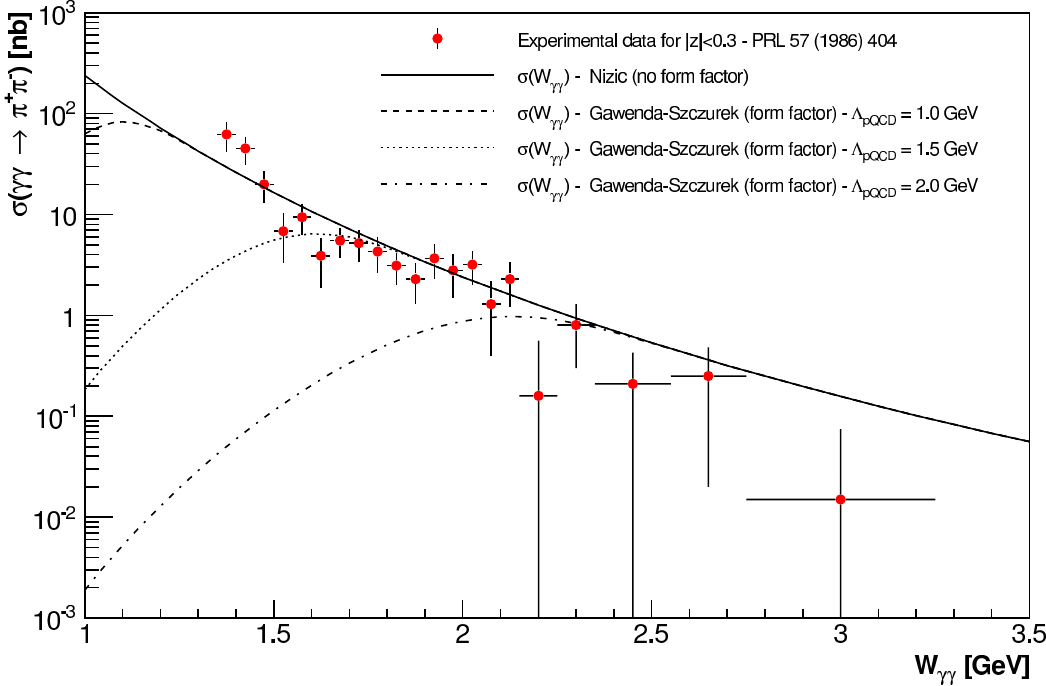} 
\par\end{centering}
\caption{Total cross section for pion pair exclusive production. Results are also computed with the pQCD contribution
suppressed by the form factor given in Eq.(\ref{e26}).}
\label{fig3} 
\end{figure}

In Fig.(\ref{fig3}) we plot the total cross section for hard exclusive two photon production of a charged pion pair.
Again our results seem to be in agreement with the existent data when calculated with the same parameters used in the
previous sections. Our results are compatible with the ones presented by Ni\v{z}i\'c \cite{nizic001}.

\begin{figure}[h]
\begin{center}
\includegraphics[scale=0.4]{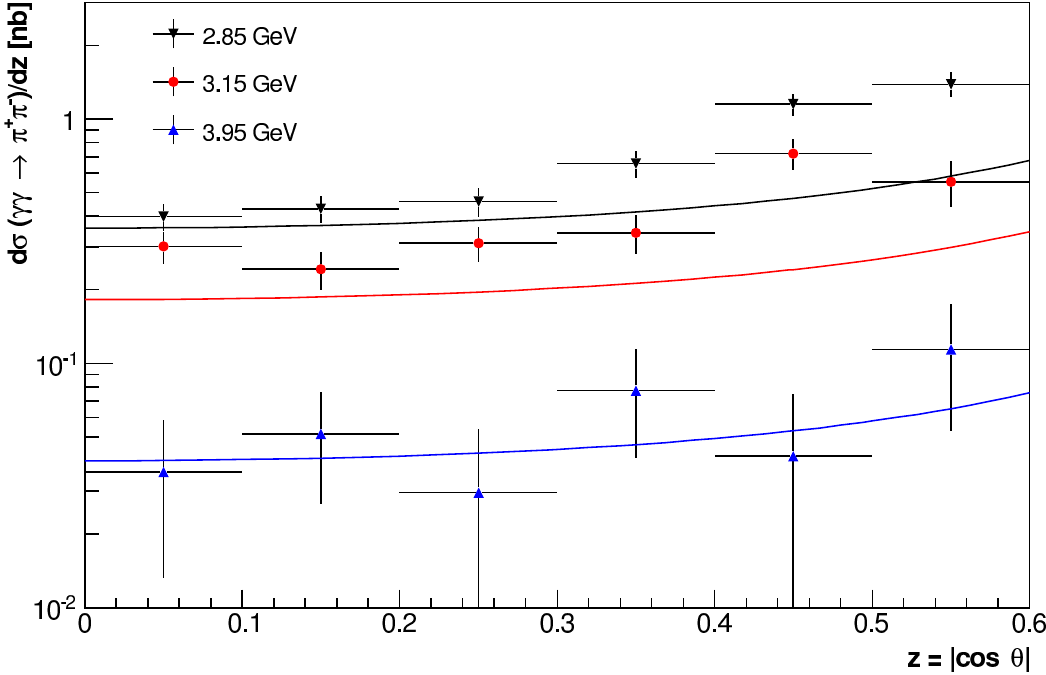} 
\end{center}
\caption{Differential cross section for pion pair exclusive production, compared with experimental data at different energies.}
\label{fig4} 
\end{figure}

Within the same approach we can compute the differential cross section for exclusive pion pair production. The existent
models, the BL one and the one of Ref.\cite{hb}, are not fully in agreement with the experimental data. This cross section
is plotted in Fig.(\ref{fig4}) and we verify that at least for large photon pair energy, where we do expect that perturbative
QCD can describe the experimental data, our calculation is consistent with the known experimental results. Unfortunately
it is still a challenge the full explanation of the experimental data within perturbative QCD, i.e. if we have already arrived
at the high energy frontier in this particular case.

\section{Discussion and conclusions}

The BaBar results for the pion transition form factor suggested many authors to propose a flat pion distribution amplitude in order to describe
the data. In Ref.\cite{luna} we proposed that only a very hard BSE solution (in momentum space) for the pion wave function can generate such
flat DA. We computed the DA as a function of this type of solution of quark self-energy, which is related to the pion wave function, and our
main intention in this work was to verify how this DA describe the experimental data. We stress that, as far as we know, only a very hard
(in momentum space) quark self-energy can lead to a natural explanation of a flat pion DA within first QCD principles.

We computed the pion transition form factor, the pion form factor and the exclusive photoproduction of charged pion pairs at high energies with the
DA determined in Section II. Following Radyushkin \cite{rad} we have assumed that QCD corrections barely affect such flat DA, however the QCD corrections
in the hard scattering amplitudes seem to be necessary for a better description of the experimental data. All quantities were computed with the
same parameters used to determine the DA, i.e. dynamical quark and gluon masses, providing a consistent picture of pions exclusive production.

In principle we may not expect that the quark self-energy, or the similar pion wave function, should follow exactly the behavior of Eq.(\ref{eq1a}),
but this is indeed one possibility that appears when the quark masses are dynamically generated \cite{natale001,natale002}, and this possibility should be
confronted with the experimental data.
However the description of the data is quite reasonable and seems to indicate that the pion wave function may be well approximated at large momentum by the
behavior of Eq.(\ref{eq1a}). Our discussion  immediately raises two important questions: will the pion DA evolve and, for very large $\mu$, will satisfy
some asymptotic behavior? The answer is yes for both. For $\mu \gg 1$ the perturbative evolution dominates and brings our pion
DA to some of the asymptotic forms discussed in Refs. \cite{rad,poly,te2,te6,roberts001,f1,f2,f3,f3l,luna,roberts002}.
However, our results depend mainly on the shape of $\varphi_{\pi}(x,\mu)$ at low scales, namely
$\mu \lesssim 1$ GeV,
and on these scales dominates the non-perturbative evolution, which is extremely slow. This can be understood by considering a particular argument due to
Radyushkin \cite{rad}: let us consider the one-loop correction $F_{\pi\gamma\gamma^{*}}^{NLO}(Q^{2})$ for the pion transition form factor, Eqs. (\ref{eqb}) and
(\ref{eqb1}),
\begin{eqnarray}
  F_{\pi\gamma\gamma^{*}}^{NLO}(Q^{2}) &=& \int_{0}^{1}dx \, \frac{\varphi_{\pi}(x,\mu)}{xQ^{2}} \left\{ 1 + \frac{4}{3} \frac{\alpha_{s}(\mu^{2})}{2\pi}
  \left[ \frac{1}{2} \ln^{2}x \right. \right. \nonumber \\
    & & - \left. \left.  \frac{x \ln x}{2\bar{x}} - \frac{9}{2}  + \left( \frac{3}{2} + \ln x  \right) \ln \left( \frac{Q^{2}}{\mu^{2}} \right)  \right]
  \right\} \nonumber \\
  &\equiv & \frac{J(Q,\mu)}{Q^{2}},
  \label{tyup001}
\end{eqnarray}
where $F_{\pi\gamma\gamma^{*}}(Q^{2}) = \frac{\sqrt{2}f_{\pi}}{3} [F_{\pi\gamma\gamma^{*}}^{LO}(Q^{2})+F_{\pi\gamma\gamma^{*}}^{NLO}(Q^{2})]$. From (\ref{eq3A}) and
(\ref{tyup001}) we have
\begin{eqnarray}
  J_{\epsilon}(Q,\mu) &=& \left( \frac{1}{\epsilon} + 2 \right) \left\{ 1 + \frac{\alpha_{s}(\mu^{2})}{3\pi} \left[ \frac{2}{\epsilon^{2}} + \frac{\pi^2}{3}
    - 9 + {\cal O}(\epsilon) \right. \right.  \nonumber \\
    & & - \left. \left. \left( \frac{2}{\epsilon} - 3 + \frac{\pi^2}{3}\epsilon + {\cal O}(\epsilon^2) \right) \ln \left( \frac{Q^{2}}{\mu^{2}} \right)
    \right] \right\} .
\label{tyup002}
\end{eqnarray}
We see that $\ln (Q^2/\mu^2)$ is very large when $Q \gg \mu$. In order to make the one-loop correction small, we can eliminate the logarithmic contribution
by adopting $Q=\mu$. We shall, however, continue to have a large one-loop correction since, in effect, the dominant term in (\ref{tyup002}) is now the term
of order $\sim 2/\epsilon^2$. In our case ($\epsilon=0.0248$) we have a huge correction $\sim 750(\alpha_{s}/\pi)$. 
In fact the dominant term can be
compensated by the logarithmic term $(2/\epsilon)\ln (Q^2/\mu^2)$ only by taking $\ln (Q^2/\mu^2)=1/\epsilon$, and this choice simply imposes
$\mu^2 = e^{-1/\epsilon}Q^2$. This relation may be rewritten as $\mu^2 = \sqrt{\bar{x}}\, Q^2$, where
\begin{eqnarray}
\ln \bar{x} \simeq \frac{\int_{0}^{1} \ln x \, \varphi_{\pi}(x;\epsilon)dx}{\int_{0}^{1}\varphi_{\pi}(x;\epsilon)dx}
\end{eqnarray}
for the case of the amplitude with almost flat DA. In the expression above $\bar{x}$ is an effective average $x$.
The optimal choice for the normalization scale $\mu$ is therefore something like
\begin{eqnarray}
\mu^2 = e^{-1/0.0248}Q^2 \sim 10^{-8}Q^2.
\label{tyup003}
\end{eqnarray}
We observe that even for the highest $Q^2$ scales reached in the BELLE and BABAR experiments, we obtain $\mu^2 \sim 4 \times 10^{-7}$ GeV$^2$. This
scale corresponds to distances much larger than the size of the pion.  
Thus, we unfortunately cannot evolve $\varphi_{\pi}(x,\mu)$ down to such tiny scale, the perturbative evolution must stop for a scale of the order of
$\mu_{0}^2 = \Lambda_{QCD}^2$. 
In other words, our flat pion DA becomes a pion DA defined at some low normalization point $\mu = \mu_{0} \sim \Lambda_{QCD}$, and below this point there is
no evolution.
In this picture we can consider the case where the evolution of $\varphi_{\pi}(x,\mu)$ is absent, i.e. we deal simply with $\varphi (x;\epsilon)$.
Hence our pion DA almost does not evolve and there is no need to specify the renormalization scale $\mu$ at which it is defined.  
It may be worth emphasizing that the argument just exposed is valid only
up to NLO correction. Choosing a special scale will not eliminate high order terms.

It is worth remarking on the fact that our pion DA, despite having an intrinsic flat
invariant (independent of the renormalization scale $\mu$) behavior across the widest range of scales, exhibits the correct UV asymptotic behavior in
the limit
$\mu \to \infty$. At this stage we have to remember that in essence the pion DA depends on the scale $\mu$ that is used to define the matrix elements of
the leading
(twist-2) local operators. Its evolution equation in kernel form is given by \cite{efr,lep}
\begin{eqnarray}
\mu \frac{d \varphi_{\pi}(x,\mu) }{d\mu} = \int_{0}^{1} V(x,y)\, \varphi_{\pi}(y,\mu) dy,
\label{rblequation}
\end{eqnarray}
where $V(x,y)$ is the evolution kernel. The general solution of (\ref{rblequation}) may be written in terms of Gegenbauer polynomials,
\begin{eqnarray}
\varphi_{\pi}(x,\mu) &=& \varphi_{\pi}^{as}(x) \left\{ 1 + \sum_{n=1}^{\infty} a_{2n}\, C_{2n}^{3/2} (2x-1) \right. \nonumber \\
& & \times \left. \left[ \ln \left( \frac{\mu^{2}}{\Lambda^{2}} \right) \right]^{-\gamma_{2n}/\beta_{0}} \right\},
\label{Gegenbauer001}
\end{eqnarray}
where $\beta_{0}$ is the first coefficient of the $\beta$ function of the QCD, and $\gamma_{2n} > 0$ is the anomalous dimension of the composite operator
(with $2n$ derivatives). As a result, in the limit $\mu \to \infty$ the pion DA acquires the asymptotic form $\varphi_{\pi}^{as}(x)$. In our case, where 
$\mu^2 \sim 10^{-8}Q^2$, the asymptotic behavior occurs on $Q^{2}$ scales far greater than the currently accessible energy regimes. We see that even in this
scenario the infrared behavior is important for short-distance quantities such as the pion DA.

It would be interesting to compare the DA calculation of Eq. (2) with the one using the Bethe-Salpeter amplitude of Ref. (18). In a
naive analysis we have verified that the form
factor calculation using the BSA, when expressions for vertex and propagators are plugled in the BSA expression, generates a similar
power
dependence on the quark self-energy and exchanged momenta. As we are using a very hard self-energy, all integrations will be quite
dependent only on
the UV asymptotic behavior and not on the IR subtleties of the calculation. Therefore we do not expect large divergences from one
result to the
other. However it is clear that a full comparison of the different calculations can motivate a more lengthy and detailed work.
A detailed comparison will be addressed in a future work.

\section*{Acknowledgments}

This research was partially supported by grant 303588/2018-7 of the Conselho Nacional de Desenvolvimento Cient\'{\i}fico e Tecnol\'{o}gico (CNPq),
by grant 2013/22079-8 of Funda\c{c}\~{a}o de Amparo \`{a} Pesquisa do Estado de S\~ao Paulo (FAPESP),
by the project INCT-FNA Proc. No. 464898/2014-5,
by the PEDECIBA program, and
by the ANII-FCE-1-126412 project. MP thanks the Universidade Federal do Rio Grande do Sul for hospitality.


\begin {thebibliography}{99}

\bibitem{babar} B. Aubert \emph{et al.} [BaBar Collaboration], Phys. Rev. D \textbf{80} (2009) 052002.

\bibitem{bl} S.J. Brodsky, G.P. Lepage, Adv. Ser. Direct. High Energy Phys. \textbf{5} (1989) 93.

\bibitem{brodsky2} G.P. Lepage, S.J. Brodsky, Phys. Rev. D \textbf{22} (1980) 2157.

\bibitem{belle} S. Uehara \textit{et al.} [Belle Collaboration], Phys. Rev. D \textbf{86} (2012) 092007.

\bibitem{rad} A.V. Radyushkin, Phys. Rev. D \textbf{80} (2009) 094009.

\bibitem{poly} M.V. Polyakov, JETP Lett. \textbf{90} (2009) 228.

\bibitem{te1} A.E. Dorokhov, JETP Lett. \textbf{92} (2010) 707.

\bibitem{te2} A.E. Dorokhov, Nucl. Phys. Proc. Suppl. \textbf{225-227} (2012) 141.

\bibitem{te3} S.V. Mikhailov, N.G. Stefanis, Nucl. Phys. B \textbf{821} (2009) 291.

\bibitem{te4} S.J. Brodsky, F.-G. Cao, G.F. de Teramond, Phys. Rev. D \textbf{84} (2011) 033001.

\bibitem{te5} E. Ruiz Arriola, W. Broniowski, Phys. Rev. D \textbf{81} (2010) 094021.

\bibitem{te6} S. Noguera, V. Vento, Eur. Phys. J. A \textbf{46} (2010) 197.

\bibitem{te7} S.S. Agaev, V.M. Braun, N. Offen, F.A. Porkert, Phys. Rev. D \textbf{83} (2011) 054020.

\bibitem{te8} P. Kroll, Eur. Phys. J. C \textbf{71} (2011) 1623.

\bibitem{te9} Y. Klopot, A. Oganesian, O. Teryaev, Phys. Rev. D \textbf{87} (2013) 036013.

\bibitem{efr} A.V. Efremov, A.V. Radyushkin, Phys. Lett. B \textbf{94} (1980) 245.

\bibitem{roberts001} C. Shi, C. Chen, L. Chang, C.D. Roberts, S.M. Schmidt, H.-S. Zong, Phys. Rev. D \textbf{92} (2015) 014035.

\bibitem{roberts002} I.C. Clo\"et, L. Chang, C.D. Roberts, S.M. Schmidt, P.C. Tandy, Phys. Rev. Lett. \textbf{111} (2013) 092001;
L. Chang, I.C. Clo\"et, C.D. Roberts, S.M. Schmidt, P.C. Tandy, Phys. Rev. Lett. \textbf{111} (2013) 141802.

\bibitem{f1} F.M. Dittes, A.V. Radyushkin, Sov. J. Nucl. Phys. \textbf{34} (1981) 293.

\bibitem{f2} I.V. Anikin, A.E. Dorokhov, L. Tomio, Phys. Lett. B \textbf{475} (2000) 361.

\bibitem{f3} E. Ruiz Arriola, W. Broniowski, Phys. Rev. D \textbf{66} (2002) 094016.

\bibitem{f3l} M. Praszalowicz, A. Rostworowski, Phys. Rev. D \textbf{64} (2001) 074003.

\bibitem{huang} X.-G. Wu, T. Huang, Phys. Rev. D \textbf{82} (2010) 034024.

\bibitem{huang2} T. Huang, T. Zhong, X.-G. Wu, Phys. Rev. D \textbf{88} (2013) 034013.

\bibitem{luna} E.G.S. Luna, A.A. Natale, J. Phys. G \textbf{42} (2015) 015003.

\bibitem{latchin} J.-H. Zhang {\it {et al.}}, Phys. Rev. D \textbf{95} (2017) 094514.

\bibitem{dor} A.E. Dorokhov, JETP Lett. \textbf{77} (2003) 63 [Pisma Zh. Eksp. Teor. Fiz. \textbf{77} (2003) 68].

\bibitem{ds} R. Delbourgo, M.D. Scadron, J. Phys. G \textbf{5} (1979) 1621.

\bibitem{BS001} J. Carbonell, V.A. Karmanov, Eur. Phys. J. A {\bf 27} (2006) 11;
J. Carbonell, V.A. Karmanov, Eur. Phys. J. A {\bf 46} (2010) 387.

\bibitem{lane} K. Lane, Phys. Rev. D \textbf{10} (1974) 2605.

\bibitem{lan} P. Langacker, Phys. Rev. Lett. \textbf{34} (1975) 1592.

\bibitem{cs} J.M. Cornwall, R.C. Shellard, Phys. Rev. D \textbf{18} (1978) 1216.

\bibitem{man} S. Mandelstam, Proc. R. Soc. A \textbf{233} (1955) 248;
C.H. Llewellyn Smith, Nuovo Cimento, A \textbf{60} (1969) 348.

\bibitem{us3} A. Doff, E.G.S. Luna, A.A. Natale, Phys. Rev. D \textbf{88} (2013) 055008.

\bibitem{poli} H.D. Politzer, Nucl. Phys. B \textbf{117} (1976) 397.

\bibitem{takeuchi} T. Takeuchi, Phys. Rev. D \textbf{40} (1989) 2697;
K.-I. Kondo, S. Shuto, K. Yamawaki, Mod. Phys. Lett. A \textbf{6} (1991) 3385.

\bibitem{us1} A. Doff, F.A. Machado, A.A. Natale, Annals of Physics \textbf{327} (2012) 1030.

\bibitem{us2} A. Doff, F.A. Machado, A.A. Natale, New. J. Phys. \textbf{14} (2012) 103043.

\bibitem{chan} L.-N. Chang, N.-P. Chang, Phys. Rev. D \textbf{29} (1984) 312;
L.-N. Chang, N.-P. Chang, Phys. Rev. Lett. \textbf{54} (1985) 2407;
N.-P.Chang, D.S. Li, Phys. Rev. D \textbf{30} (1984) 790.

\bibitem{mon} J.C. Montero, A.A. Natale, V. Pleitez, S.F. Novaes, Phys. Lett. B \textbf{161} (1985) 151.

\bibitem{natale001} A.C. Aguilar, A. Doff, A.A. Natale, Phys. Rev. D {\bf 97} (2018) 115035.

\bibitem{natale002} A. Doff, A.A. Natale, Eur. Phys. J. C {\bf 78} (2018) 872.

\bibitem{cornwall} J.M. Cornwall, Phys. Rev. D \textbf{26} (1982) 1453.

\bibitem{us4} C.A.S. Bahia, M. Broilo, E.G.S. Luna, Phys. Rev. D \textbf{92} (2015) 074039;
D.A. Fagundes, E.G.S. Luna, M.J. Menon, A.A. Natale, Nucl. Phys. A \textbf{886} (2012) 48;
E.G.S. Luna, A. A. Natale, A.L. dos Santos, Phys. Lett. B \textbf{698} (2011) 52;
E.G.S. Luna, Phys. Lett. B \textbf{641} (2006) 171;
E.G.S. Luna, A.F. Martini, M.J. Menon, A. Mihara, A.A. Natale, Phys. Rev. D \textbf{72} (2005) 034019.

\bibitem{ap} A.C. Aguilar, J. Papavassiliou, JHEP \textbf{0612} (2006) 12;
A.C. Aguilar, J. Papavassiliou, Eur. Phys. J. A \textbf{31} (2007) 742.

\bibitem{ap1} A.C. Aguilar, D. Binosi, J. Papavassiliou, Front. Phys. China {\bf 11} (2016) 111203.

\bibitem{che} V.L. Chernyak, A.R. Zhitnitsky, JETP Lett. \textbf{25} (1977) 510.

\bibitem{lep} G.P. Lepage, S.J. Brodsky, Phys. Lett. B \textbf{87} (1979) 359.

\bibitem{brodsky} S.J. Brodsky, G.P. Lepage, Phys. Rev. D \textbf{24} (1981) 1808.

\bibitem{aguila} F. del Aguila, M.K. Chase, Nucl. Phys. B \textbf{193} (1981) 517.

\bibitem{braaten} E. Braaten, Phys. Rev. D \textbf{28} (1983) 524.

\bibitem{halnat} M.B. Gay Ducati, F. Halzen, A.A. Natale, Phys. Rev. D \textbf{48} (1993) 2324.

\bibitem {aguinat} A.C. Aguilar, A. Mihara, A.A. Natale, Phys. Rev. D \textbf{65} (2002) 054011.

\bibitem{ji} C.-R. Ji, F. Amiri, Phys. Rev. D {\bf 42} (1990) 3764;
S.J. Brodsky, C.-R. Ji, A. Pang, D.G. Robertson, Phys. Rev. D {\bf 57} (1998) 245.

\bibitem{brodsky3} S.J. Brodsky, SLAC-PUB-7604, arXiv:hep-ph/9708345;
  S.J. Brodsky, arXiv:hep-ph/9710288.

\bibitem{dalley001} S. Dalley, B. van de Sande, Phys. Rev. D {\bf 67} (2003) 114507.

\bibitem{agui} A.C. Aguilar, A.A. Natale, JHEP \textbf{0408} (2004) 057.

\bibitem{papa} J. Papavassiliou, J. Phys.: Conf. Ser. \textbf{631} (2015) 012006. 

\bibitem{yeh} Tsung-Wen Yeh, Phys. Rev. D \textbf{65} (2002) 074016.

\bibitem{bebek01} C.J. Bebek \textit{et al.}, Phys. Rev. Lett. {\bf 37} (1976) 1525;
C.J. Bebek \textit{et al.}, Phys. Rev. D {\bf 17} (1978) 1693.

\bibitem{volmer01} J. Volmer \textit{et al.}, Phys. Rev. Lett. 86 (2001) 1713.

\bibitem{horn01} T. Horn \textit{et al.}, Phys. Rev. Lett. 97 (2006) 192001.

\bibitem{tadevosyan01} V. Tadevosyan \textit{et al.}, Phys. Rev. C 75 (2007) 055205.

\bibitem{gs} M.K.-Gawenda, A. Szczurek, Phys. Rev. C \textbf{87} (2013) 054908.

\bibitem{nizic001} B. Ni\v{z}i\'c, Phys. Rev. D {\bf 35} (1987) 80.  

\bibitem{hb} M. Diehl, P. Kroll, C. Vogt, Phys. Lett. B \textbf{532} (2002) 99; 
M. Diehl, P. Kroll, Phys. Lett. B \textbf{683} (2010) 165.

\end {thebibliography}
\end{document}